\documentclass[10pt,twoside,a4paper,twocolumn]{article}

\usepackage[utf8]{inputenc}
\usepackage{graphicx}
\usepackage{dcolumn}
\usepackage{bm}
\usepackage{authblk}
\usepackage{amssymb}
\usepackage{amsmath}
\usepackage{flushend}
\usepackage{placeins}  
\usepackage{hyperref}
\usepackage{abstract}
\usepackage{arydshln}  
\usepackage{color} 
\usepackage{cuted}
\usepackage[version=4]{mhchem}
\usepackage[a4paper, portrait, margin=21mm]{geometry}  

\renewcommand{\L}{\ensuremath{\mathcal{L}}}
\newcommand{\cov}{\ensuremath{\text{cov}}}

\newcommand{\C}{\ensuremath{\mathcal{C}}}
\newcommand{\constrained}{\emph{Constrained}}
\newcommand{\unconstrained}{\emph{Unconstrained}}
\newcommand{\excluded}{\emph{Excluded}}
\newcommand{\conventional}{\emph{Conventional}}
\newcommand\sbullet[1][.5]{\mathbin{\vcenter{\hbox{\scalebox{#1}{$\bullet$}}}}}

\makeatletter
\renewcommand*{\@fnsymbol}[1]{\ensuremath{\ifcase#1\or *\or \dagger\or \ddagger
\mathsection\or \mathparagraph\or \|\or **\or \dagger\dagger
\or \ddagger\ddagger \else\@ctrerr\fi}}
\makeatother

\date{\vspace{-5ex}}

\begin{document}

\title{Improved precision and accuracy of electron energy-loss spectroscopy quantification via fine structure fitting with constrained optimization}

\author[,a,b]{Daen~Jannis \thanks{Now at Thermo Fisher Scientific, Achtseweg Noord 5, 5651 GG Eindhoven, Netherlands.}}%
\author[,c]{Wouter~Van~den~Broek \thanks{Email: \url{wouter.vandenbroek@thermofisher.com}}}%
\author[a,b,d]{Zezhong~Zhang}%
\author[a,b]{Sandra~Van~Aert}%
\author[a,b]{Jo~Verbeeck}%

\affil[a]{Electron Microscopy for Materials Research (EMAT), University of Antwerp, Groenenborgerlaan 171, 2020 Antwerp, Belgium.}
\affil[b]{Nanolab center of excellence, University of Antwerp,  Groenenborgerlaan 171, 2020~Antwerp,~Belgium.}
\affil[c]{Thermo Fisher Scientific, Achtseweg Noord 5, 5651 GG Eindhoven, Netherlands.}
\affil[d]{Department of Materials, University of Oxford, 16~Parks~Road, Oxford~OX1~3PH,~United~Kingdom.}


\twocolumn[%
{\footnotesize\copyright \ CC-BY-NC-ND 4.0: \url{https://creativecommons.org/licenses/by-nc-nd/4.0/}}%
\maketitle%
\textbf{Abstract} \quad By working out the Bethe sum rule, a boundary condition that takes the form of a linear equality is derived for the fine structure observed in ionization edges present in electron energy-loss spectra. This condition is subsequently used as a constraint in the estimation process of the elemental abundances, demonstrating starkly improved precision and accuracy and reduced sensitivity to the number of model parameters.  Furthermore, the fine structure is reliably extracted from the spectra in an automated way, thus providing critical information on the sample's electronic properties that is hard or impossible to obtain otherwise.  Since this approach allows dispensing with the need for user-provided input, a potential source of bias is prevented.\\

\textbf{Keywords} \quad Electron energy-loss spectroscopy; Elemental abundance; Fine structure; Constrained optimization; Bethe sum rule; Quadratic programming; Cram\'er-Rao lower bound\\
]\saythanks

\section{Introduction}

Electron energy-loss spectroscopy (EELS) is a powerful technique which measures the energy-loss distribution of high-energy electrons after interaction with matter. The energy-loss reflects the inelastic scattering probability for the electron, providing information about the elemental abundances and electronic structure of the material of the sample. The technique is frequently utilized in combination with scanning transmission electron microscopy (STEM) enabling elemental and electronic structure mapping on the atomic scale. 

Even though EELS has been around for multiple decades, quantification of the data remains challenging and heavily dependent on user input, thus risking (unwitting) user bias. For example, in the standard background subtraction method presented in \cite{egerton1996_44}, the user chooses a suitable pre-edge background estimation window, and the result is quite sensitive to this choice as pointed out by \cite{egerton2002ultram, liu1987jom, su1993prb}, among others. Furthermore, different choices of background functions are provided, thus further increasing user influence.

To relieve such issues a statistical model-based approach was developed in the 2000s by \cite{verbeeck_2004}.  It aims at retrieving elemental concentrations by fitting a model to the experimental data. This model initially consisted of a power-law background, energy-dependent atomic cross sections for each recorded edge, and a convolution with a zero-loss peak recording to account for multiple scattering. Hence, the model is a function of various parameters, in particular the elemental abundances. 

An important benefit of statistical parameter estimation theory is the availability of the so-called Cram\'er-Rao bound (CRB), which defines a lower bound on the variance of the estimated model parameters \cite{vandenbos2001hawkes, verbeeck2008ultram} in the absence of bias. This concept can thus be used to derive an expression for the highest attainable precision with which, for example, the elemental concentrations can be estimated. The CRB can be used to either design an optimal experimental setup---so-called statistical experimental design, see for example \cite{dendekker2013ultram, vandenbroek2019ieee}---, verify if the estimator attains its full potential at best precision, or as the error estimates of the estimation procedure.

Realistic EEL spectra feature fine structure: fine details in the spectra that are due to the atoms' chemical environment and are hence not accounted for by the more smooth atomic cross section models from above.  Fitting procedures typically deal with this fact by excluding from the fit the energy-loss near-edge structure (ELNES) region, where fine structure effects are strongest and which extends 30~to~50~eV from the edge onset;  \cite{rez2004, egerton1996_38}.  As we will show, this is not optimal in terms of precision of the elemental abundance estimates because it removes the region with the strongest signal-to-background ratio, and introduces a bias because the fine structure effects linger on far beyond the typically excluded energy range.

Hence, the model was improved further by \cite{verbeeck2006ultram} by adding a multiplicative fine structure to the atomic cross sections. This model consists of a heuristic piece-wise linear function, whose coefficients have to be independently estimated.  Under some mild assumptions the multiplicative fine structure approximately equals the unoccupied density of states in a crystal relative to the density of states of a free atom. Access to the fine structure is hence important as it provides critical information on the elements' electronic surroundings.

\cite{cueva2012} replaced the single power-law background with variable exponent with a linear combination of power-laws with fixed exponents.  By doing away with the variable exponent, the only non-linear parameter is removed from the model, and the estimation process has a guaranteed minimum and becomes faster and more robust. To ensure that the fitted background is non-negative, descending and convex over the entire energy range, \cite{vandenbroek2023ultram} formulated linear inequality constraints that were imposed through quadratic programming. The result is a background model that outperforms the conventional power-law approach.

In this work, an additive fine structure model is added to the atomic cross sections. It has the advantage of allowing the formulation of a physically meaningful linear equality constraint on the fine structure that is derived from the Bethe sum rule.  By imposing said constraint with quadratic programming, the fit is more robust, and has improved precision. Compared to conventional, non-model-based, estimation procedures, the accuracy improves as well.

The proposed technique leads to a starkly reduced need for user input, and hence improved objectivity of the results. While conventional EELS methods need an initial visual inspection of the spectra to decide on parameters important for subsequent processing---for instance the width of the background window or the excluded fine structure range mentioned above---, this novel method only needs information on which elements are present, and can set up the rest of the estimation problem autonomously from the experimental settings available in the spectra's meta-data. 

The paper is organized as follows.  In Sec.~\ref{sec:theory} the linear equality constraint on the fine structure is derived; Sec.~\ref{sec:met} presents a treatment of model-based EELS, quadratic programming and the Cram\'er-Rao lower bound for constrained estimation problems; in Sec.~\ref{sec:res} the performance of the constrained optimization approach is compared to three other common approaches with the aid of simulations and experiments, and the influence on the elemental maps of a \ce{TbScO3} sample is illustrated, as is the estimation of the fine structure of \ce{Si} versus \ce{SiO2}; in Sec.~\ref{sec:dis} alternatives to the equality constraint and the bias of the method are discussed; finally, in Sec.~\ref{sec:con} the conclusions are drawn.

\section{Theory}
\label{sec:theory}

In Appendix~\ref{app:theory} it is shown that
\begin{equation} \label{eq:cte}
    \int_0^{\infty} \frac{E \sigma_{in}(E)}{\log \Big(1+\frac{\beta^2} {\theta_E(E)^2}\Big)} dE = \text{constant.}
\end{equation}
This expression states that the integral of the inelastic cross section, $\sigma_{in}$, with weights depending on energy-loss, $E$, collection angle, $\beta$, and characteristic inelastic scattering angle, $\theta_E$, is constant and holds for any shape of the inelastic cross section dictated by the chemical environment of the atom.  Eq. \ref{eq:cte} hence does not depend on the atom's surroundings, be it vacuum or any type of chemical bonds.

The derivation starts from the Bethe sum rule and uses the following approximations:
\begin{enumerate}
   \item non-relativistic Schr\"odinger equation;
   \item the contribution from occupied states is independent of the atomic environment; 
   \item the particular shell electron wavefunctions are orthogonal to all other electrons; 
   \item the collection angle is small (dipole approximation), leading to a constant generalized oscillator strength (GOS),  \cite{crozier1989ultram}.
\end{enumerate}

We include additive fine structure through some set of basis functions, $g_i(E)$, defined on the energy range were the fine structure occurs, 
\begin{equation} \label{eq:fine}
    \sigma_{in}(E) = \sum_{i=1}^{m} b_i g_i(E) + \sigma_A(E),
\end{equation}
with $b_i$ the fine structure coefficients that need estimation and $\sigma_A$ the atomic inelastic cross section.  While the physical interpretation of the additive model is not immediately obvious, we show in Appendix \ref{app:1} the mathematical equivalency with the multiplicative form in \cite{verbeeck2006ultram}, which in turn has a clear interpretation as the approximate unoccupied density of states in a crystal relative to the density of states of the free atom.

As Eq.~\ref{eq:cte} holds for all physical inelastic cross sections, we use it to compare $\sigma_A$ and $\sigma_{in}$ from Eq.~\ref{eq:fine}, leading to the equality,
\begin{multline}
    \int_0^{\infty} \frac{E \sigma_{A}(E)}{\log \Big(1+\frac{\beta^2} {\theta_E(E)^2}\Big)} dE \\
     = \int_0^{\infty} \frac{E \Big(\sigma_{A}(E) + \sum b_i g_i(E) \Big)dE}{\log \Big(1+\frac{\beta^2} {\theta_E(E)^2}\Big)} dE.
\end{multline}

Next, the integral is broken into two parts. The first region [0, $E_{max}$] includes most of the fine structure while the second region [$E_{max}$, $\infty$) has negligible fine structure contributions and is therefore almost identical to the atomic cross section. In practice $E_{max}$ must be chosen sufficiently large, but lie below the onset of the closest following edge, typically ranges of 100-500~eV yield good results.

The cross sections $\sigma_A$ and $\sigma_{in}$ have equal integral over the second region, [$E_{max}$, $\infty$), hence only the contribution of [0, $E_{max}$] remains and the integral simplifies to
\begin{equation}\label{eq:constrain}
\sum_i b_i \int_0^{E_{max}} \frac{E g_i(E)}{\log \Big(1+\frac{\beta^2} {\theta_E(E)^2}\Big)} dE = 0.
\end{equation}
The integrals in the summands depend only on known parameters so that they can be calculated off-line. This means that Eq. \ref{eq:constrain} constitutes a linear equality constraint on the fine structure coefficients, $b_i$, that in this work is imposed on the fitting solution through quadratic programming.  

Note that in this treatment the atomic cross section $\sigma_A$ contains the contribution of excitations to unoccupied discrete bounded states and continuum states. The fine structure term can be physically interpreted as a \emph{redistribution}, which is why the Bethe sum rule bounds its integral to zero through Eq.~\ref{eq:constrain}.

Although the derivation so far assumed plane wave illumination, also convergent probes can be treated; see Appendix \ref{app:2}.

\begin{figure*}[t]
\centering
\includegraphics[width=0.99\textwidth]{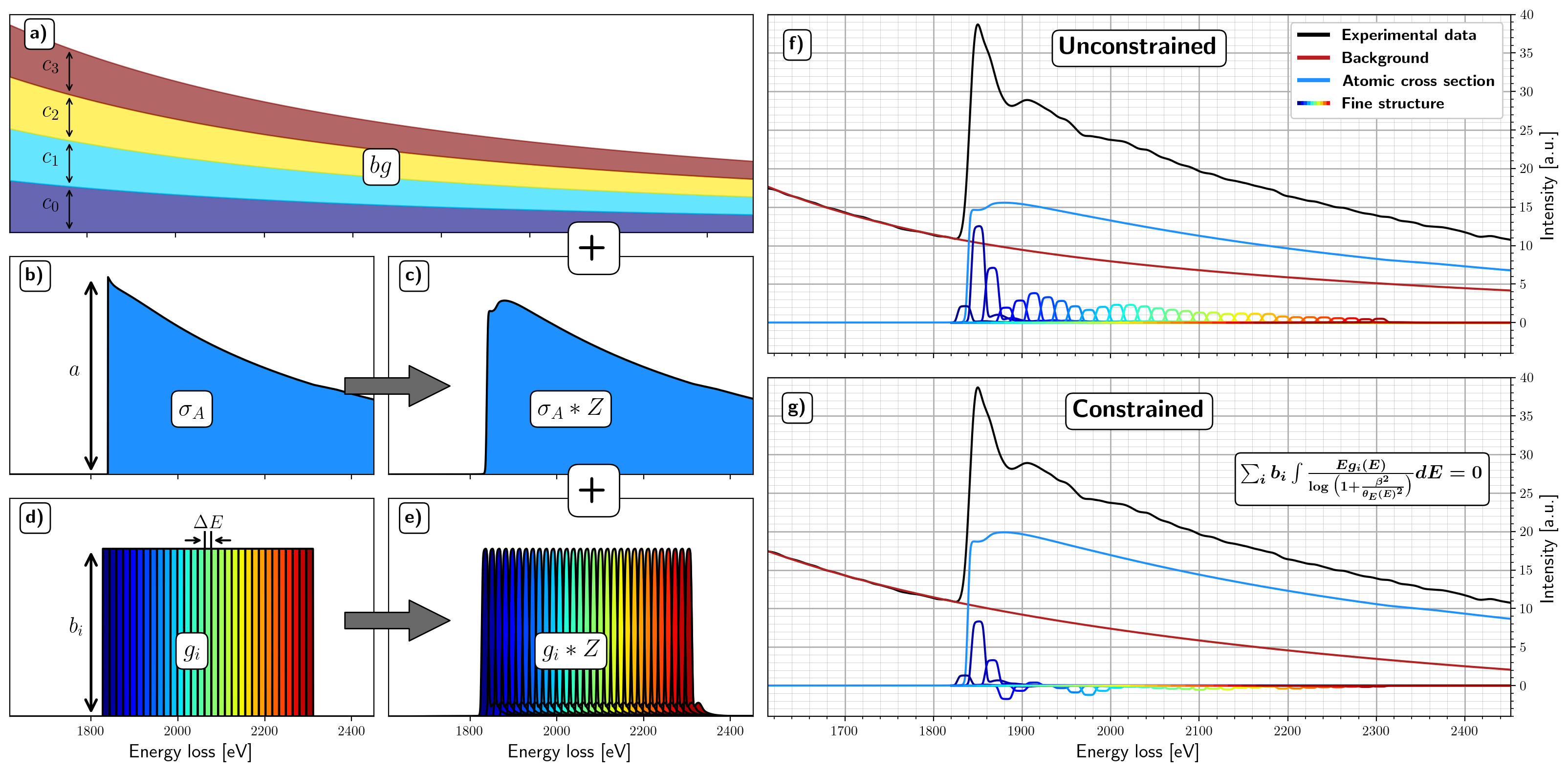}
\caption{Graphical representation of the model-based approach used in this work. (a) The background model as described by Eq.~\ref{eq:bg} and consisting of four terms. (b) The atomic cross section of the \ce{Si} K-edge. (c) The cross section convolved with the low-loss signal. (d) The fine structure components which in this case are multiple rectangular functions described by Eq~\ref{eq:finestruc} and (e) shows the convolved result. (f) An experimental \ce{Si} K-edge signal is shown in black. The resulting fit for the different components (a,c,e) is shown as well; no constraints are imposed on the fine structure. (g) Similar to (f), but the equality constraints of Eq~\ref{eq:constrain} are applied to the fine structure. Note that for visualization purposes the experimental edge in (f,g) was broadened by 8~eV to decrease the required number of fine structure functions.
\label{fig:showing}}
\end{figure*}

\section{Methods}
\label{sec:met}

\subsection{Model-based Methodology}
\label{sec:modbasmet}

In this section a description of the entire model is provided.  The intensity $I$ is given as,
\begin{multline}
    I(E) = bg(E) \ + \\ 
    Z(E) \otimes  \sum_{k=0}^n \left(\sum_{i=0}^{m} b_{i,k} g_{i,k}(E) + a_k \sigma_{A,k}(E) \right), \label{eq:model}
\end{multline}
with $bg$ the background, $Z$ the zero-loss peak, $\otimes$ indicating a convolution, the index $k$ runs over the edges present in the spectrum, $i$ enumerates the fine structure basis functions, $g_{i,k}$, while the parameter $a_k$ denotes the elemental abundances and $\sigma_{A,k}$ is the inelastic atomic cross section of edge $k$. In the remainder of this section, the different terms of this equation are worked out. 

The background signal $bg(E)$ is modeled as
\begin{equation}\label{eq:bg}
        bg(E) = \sum_{j=1}^4 c_j E^{-r_{j}},
\end{equation}
where the exponents $r_j$ equal $1$, $2.33$, $3.67$ and $5$.  Conventionally, the background is modeled with a single power-law, whose exponent needs fitting as well, making for a non-linear estimation process. Adoption of the current background model makes the whole model linear, thus enabling a linear weighted least squares fit with guaranteed minimum, and improved speed and robustness. Lastly, in \cite{vandenbroek2023ultram}, it is demonstrated that this background model describes experimental backgrounds better than the conventional single power-law. An example is shown in Fig.~\ref{fig:showing}(a). 
    
The atomic cross sections, $\sigma_A$, are calculated with Eq.~\ref{eq:gos_to_cross} using the GOS from \cite{zhang2024zenodo}. As an example, the \ce{Si} K-edge is shown in Fig.~\ref{fig:showing}(b). For most users the pre-factors to the atomic cross sections, the parameters $a_k$, are of most importance as these reflect the elemental abundances. 

The fine structure basis functions, $g(E)$, are chosen as top-hats, although other choices are valid as well,
\begin{equation}\label{eq:finestruc}
        g_{i,k} = H(E-E_i) - H(E - (E_i + \Delta E)),
\end{equation}
where $H$ is the Heavyside step function. The value for $E_i$ follows the recursive relation: $E_{i} = E_{i-1} + \Delta E$, and the initial value $E_0$ is set to the known onset energy of the $k^{\text{th}}$ edge. $\Delta E$ determines the resolution and is set to a similar value as the width of the zero-loss peak. The number of basis functions, $m$, is chosen such as to make the fine structure span an energy range of maximum $500$~eV, or to end $10$~eV ahead of the next edge. In this way $\Delta E$ and $m$ are determined automatically and no end-user input is needed.

A convolution of fine structure and atomic cross section with the low-loss signal, $Z(E)$, is required to account for multiple scattering. The background is not convolved to reduce artifacts related to the Fourier transform's periodic boundary conditions; see \cite{verbeeck_2004} for an in-depth discussion.

The model-based approach lends itself well to an analysis of bias and precision of the estimated parameters, here applied to elemental abundances in particular. The bias is defined as the deviation of the average estimate of a parameter from its true value; it is hence related to accuracy, as high accuracy implies low bias. Precision is defined as the standard deviation of the estimates of a parameter. A good estimate is said to have a high precision, which corresponds to a low standard deviation.

\subsection{Quadratic programming}
\label{sec:quapro}

Quadratic programming (QP) provides the minimizer of a multivariate quadratic function, subject to linear equality and inequality constraints; see \cite{goldfarb1983, nocedal1999}.  

The criterion being minimized here is,
\begin{eqnarray}
    \sum_i \frac{1}{W_i} \left( I(E_i) - J_i \right)^2, \label{eq:weightedleastsquares}
\end{eqnarray}
with $I(E_i)$ the model from Eq.~\ref{eq:model} evaluated in energy bin $i$, $J_i$ the measured intensity in said bin, and $W_i$ the weight obtained from a preliminary background fit to the recorded spectrum.  This expression, in turn, can be reformulated as a QP, as has been described in full detail by \cite{vandenbroek2023ultram} in the context of model-based EELS.  The respective equality and inequality constraints are described in the following two paragraphs.

We enforce non-negativity of elemental abundances by demanding that the parameters $a_k$ in Eq. \ref{eq:model} be positive or zero. Furthermore, the so-called `sufficient' inequality constraints on the parameters $c_j$ in the background are taken from \cite{vandenbroek2023ultram} to ensure a non-negative, monotonic decreasing and convex background.

Adherence to the Bethe sum rule is imposed by providing to the QP the linear constraint of Eq. \ref{eq:constrain} on the fine structure coefficients $b$ in Eq. \ref{eq:model}, using the Quadprog Python package by \cite{quadprog}.

In Fig.~\ref{fig:showing}, the fitting procedure is illustrated on an experimental Si K-edge.  The unconstrained fit in Fig.~\ref{fig:showing}(f) results in all-positive fine structure components, while the constrained fit in Fig.~\ref{fig:showing}(g), yields a more physical result.  Note the severe impact on the estimated elemental abundances, visible as a notably different amplitude of the  atomic cross sections.

The all-positive result of the unconstrained fit illustrates the problem's initial ill-conditionedness, as different noise realizations could just as well produce all-negative results.  The physically valid result of the constrained fit demonstrates the relief of the ill-conditionedness provided by the constraint.

\subsection{Cram\'er-Rao lower bound}
\label{sec:crlb}

The Cram\'er-Rao lower bound (CRB) is a lower bound on the variance of the estimated model parameters in the absence of bias, see \cite{rao1945bul, cramer1946book, lehmann2006book}.   For a sufficient number of measurements the precision of the maximum likelihood estimator attains the CRB.  

In this work, the CRB is used to investigate the influence of the number of parameters in the fine structure model, and to verify that the precisions on the elemental concentrations estimated with the investigated fitting methods approach their theoretical lower bound.

In Appendix \ref{app:crlb} the precision of the maximum likelihood estimator with linear equality constraints is derived under assumption of Poisson noise, resulting in the following expression for the covariance matrix,
\begin{eqnarray}
    \cov_p = 
    \left . \begin{pmatrix}
     & & & & & \vline & w_1 \\
     & & & & & \vline &  \\
     \multicolumn{5}{c}{\scalebox{1.41}{$\sum_i \frac{1}{f_i}  \frac{\partial f_i}{\partial p_j} \frac{\partial f_i}{\partial p_k}$}} & \vline & \vdots \\
     & & & & & \vline &  \\
     & & & & &  \vline &  w_N \\
     \hline
     w_1 &  & \hdots & & w_N & \vline & 0
    \end{pmatrix}^{-1} \right\vert_{1 \cdots N, 1 \cdots N} \label{eq:crlb_short}
\end{eqnarray} 
Here, $f$ is the model for the measurements, the index $i$ runs over the energy bins, and $p$ is the vector containing all $N$ model parameters. The parameters $w$ represent the weights in the linear constraint and equal zero whenever the associated model parameter is not involved in the constraint.

Applied to the problem at hand, namely the model of Eq.~\ref{eq:model}, $f_i$ equals $I_i$; the parameters $a_k$ and $b_{i,k}$, and $c_j$ from Eq.~\ref{eq:bg} constitute the vector $p$; and the non-zero entries of $w$ equal the value of the integral in the summands of Eq.~\ref{eq:constrain}.

In the absence of constraints Eq.~\ref{eq:crlb_short} reduces to the conventional, constraint-free form for the CRB as given in Eq. \ref{eq:crlb}.  This fact combined with the statement by \cite{moore2010thesis} that ``... the asymptotic variance of the [maximum likelihood estimator] equaling the CRB lends credence to the claim that the asymptotic variance of the [constrained maximum likelihood estimator] should equal ... the CRB under equality constraints ...'', indicates that Appendix~\ref{app:crlb} presents an alternative derivation for the CRB under Poisson noise that can handle linear equality conditions.  For these reasons, we denote the expression in Eq.~\ref{eq:crlb_short} as a CRB for constrained estimation problems, and refer to it accordingly in the remainder of this paper.

Linear \emph{inequality} constraints, too, play an important role in modelling the background and the elemental abundances, and including them into the CRB in a formal way is not obvious.  We posit that when the estimates lie far from the boundaries the inequality constraints define in search space, so that said constraints are hardly ever active, their influence on the lower bound is negligible.  This is accomplished by an appropriate setup of the model for the simulation studies in Sec.~\ref{sec:simstu}: the background exactly matches a single term in Eq.~\ref{eq:bg} and the elemental abundance is set high enough for the rate of negative estimates to be negligible.

In this work a weighted linear least squares fitting is used which, despite being a good approximation to the maximum likelihood estimator, is expected to yield a precision slightly larger than dictated by the CRB.  This is indeed born out in Sec. \ref{sec:simstu}.

\section{Results}
\label{sec:res}

This Section features four different optimization methods that are listed here for reference.

\begin{itemize}
    \item \constrained. The model is given by Eq.~\ref{eq:model}, with all components described in Sec.~\ref{sec:modbasmet}.  Furthermore, all constraints discussed in Sec.~\ref{sec:quapro} apply: non-negativity, descent and convexity of the background; non-negativity of the elemental abundances; and the linear equality constraint on the additive fine structure.
    
    \item \unconstrained. The model is given by Eq.~\ref{eq:model}, and with the exception of the linear equality constraint on the fine structure, the same constraints as in the method \constrained \ apply; this name thus refers to the treatment of the fine structure specifically. The fine structure energy window is the same as that of \constrained, as is $\Delta E$ and the number of basis functions $g$.
    
    \item \excluded. The model does not include fine structure, but the other components are present.  The same constraints apply, except of course that of the fine structure. An energy range immediately following the edge onset, the so-called fine structure window, is excluded form the fit, as this is common practice to mitigate the unaccounted-for effects in the ELNES region where fine structure effects are strongest.
    
    \item \conventional. This is the usual background subtraction method where a single power-law background is fitted in a pre-edge window, extrapolated under the edge and then subtracted. Next, the signal is integrated over a post-edge window and the result compared to the theoretical cross section (which is convoluted with the low loss) to get the elemental abundance estimate. The pre- and post-edge window widths are difficult to standardize, as they depend on preceding and succeeding edge onsets. This necessitates heavy user involvement which in turn opens the door for user bias. In this work, we have chosen these parameters to the best of our abilities to aim for as-good-as-possible results. 
\end{itemize}
The optimization of the methods \constrained, \unconstrained \ and \excluded \ is carried out with quadratic programming.

The \conventional \ method is not underpinned by the linear least squares optimization that enables the other three methods, as such a quantitative comparison to these other methods is cumbersome.  Hence, \conventional \ is only included as an example for the elemental maps in Section \ref{sec:elmatb}.

\begin{table}
\centering
\begin{tabular}{ |c|c|c|c| } 
 \hline
   $\Delta E$ [eV] & \# unknowns & $\sigma_{unc}$ [\%] & $\sigma_{con}$ [\%] \\ 
 \hline
  0.5 & 270 & 16.6 & 2.7 \\
  1.0 & 137 & 11.7 & 2.7 \\
  2.0 & 71 & 8.3 & 2.7 \\
  4.0 & 37 & 5.9 & 2.6 \\
  8.0 & 21 & 4.2 & 2.5 \\
  16.0 & 12 & 3.0 & 2.3 \\
  \hline
\end{tabular}
 \caption{The relative errors, $\sigma_{unc}$ and $\sigma_{con}$, of the abundance estimates of \unconstrained \ and \constrained, respectively, have been computed with the CRB-formalism for various values of $\Delta E$ and associated number of unknowns. The relative error of \excluded \ (with a fine structure window of 50~eV) came out as $2.9$\%.}
\label{tab:crlb}
\end{table}

\subsection{Cram\'er-Rao Lower bound investigation}
\label{sec:crlbinv}

We use the CRB to investigate the influence the linear equality constraints on the fine structure have on the precision of the elemental abundance estimates of \unconstrained, \constrained \ and \excluded.

Consider the toy model,
\begin{eqnarray}
  f_i =
  \begin{cases}
      \sbullet \ \ c E_i^{-r}, \ \ \ \ \text{if } E_i < E_O \\
      \begin{aligned}
      \sbullet \ \ a E_i^{-2.5} & + \sum_j b_j g_j(E_i) \\
      & + c E_i^{-r} \ \ \ \ \text{if } E_i \geq E_O.
      \end{aligned}
  \end{cases}
\end{eqnarray}
where the index $i$ runs over the $n$ energy bins, $cE^{-r}$ describes a background, $E_O$ is the ionization energy and hence denotes the onset of the associated ionization edge, which in turn is modeled as $aE^{-2.5}$, and the last term describes the additive fine structure with $b_j$ the coefficients and $g_j$ the basis functions from Eq.~\ref{eq:finestruc}.  Furthermore, we have the equality constraint
\begin{eqnarray}
    \sum_j w_j b_j = 0,
\end{eqnarray}
acting on the fine structure, with $w_j$ the values of the weights.

The following values were set for the model parameters: $r = 3$, $a = 1.12 \times 10^9$, $b_j = 0$ for all $j$, $c = 1.25 \times 10^{11}$, $E_O = 500$~eV, the dispersion is $0.125$~eV, and the energy axis runs from $433$~to~$633$~eV.  These choices yield an expected electron count of $1000$ at the edge onset, and a signal-to-background-ratio of $0.2$.

The conventional power-law background with variable exponent is used to impose non-negativity, descent and convexity through the model instead of through inequality constraints, and the signal-to-background-ratio is high enough that a negligible number of negative abundance estimates are expected.  As a result, the CRB with and without fine structure equality constraints holds for the \constrained \ and \unconstrained \ methods, respectively, while the CRB of the model without fine structure and exclusion of a post-edge window, holds for \excluded.

$\Delta E$ form Eq.~\ref{eq:finestruc} was varied from $0.5$~to~$16$~eV in six steps, and the weights $w_j$ were set accordingly to the energy value on which the respective basis function $g_j$ is centered. For each $\Delta E$ the CRBs of the elemental abundance, $a$, for \unconstrained \ and \constrained \ were calculated. The CRB for \excluded \ is computed too, but is independent of $\Delta E$, as its model does not contain fine structure. The results are summarized in Table~\ref{tab:crlb}. The fine structure windows of \unconstrained \ and \constrained \ are set to $133$~eV, and that of \excluded \ to 50~eV.

The relative error of \constrained \ is always lower than that of \unconstrained, and depends only very weakly on the number of unknowns: it stays virtually constant while the unknowns vary with a factor of $24$. \excluded \ yields an error that, although it mostly improves on \unconstrained, is always larger than that of \constrained.  The relatively large error of \excluded, considering its low number of fitting parameters, is attributed to it excluding the most intense and lowest-noise part of the spectrum from the fitting procedure.  Furthermore, it is shown in Sec.~\ref{sec:simstu} that in general \excluded \ yields biased results.

\begin{figure*}
\centering
\includegraphics[width=0.99\textwidth]{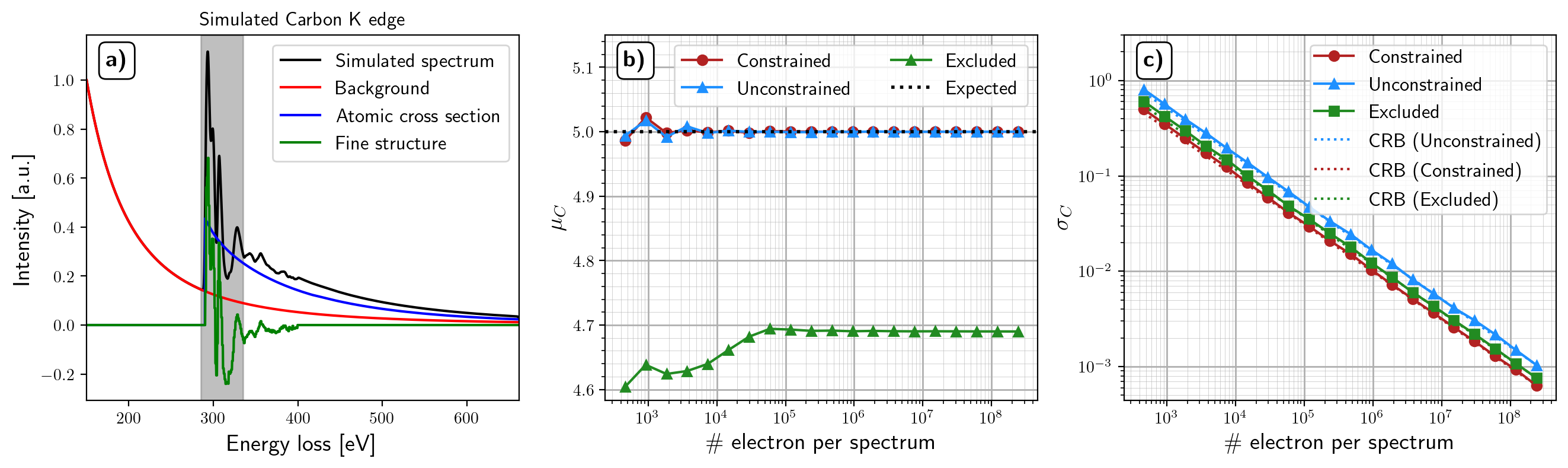}
\caption{(a) Simulated EEL spectrum with the background, atomic cross section and fine structure. The fine structure is modelled in such a way that Eq.~\ref{eq:constrain} holds. The 50~eV~wide gray area indicates the fine structure window for \excluded. (b) The average value of carbon content as a function of the number of electrons for the three methods. (c) Standard deviation and CRB on the carbon quantification as a function of the number of electrons.  
\label{fig:simulated}}
\end{figure*}

\begin{figure}
\centering
\includegraphics[width=0.99\linewidth]{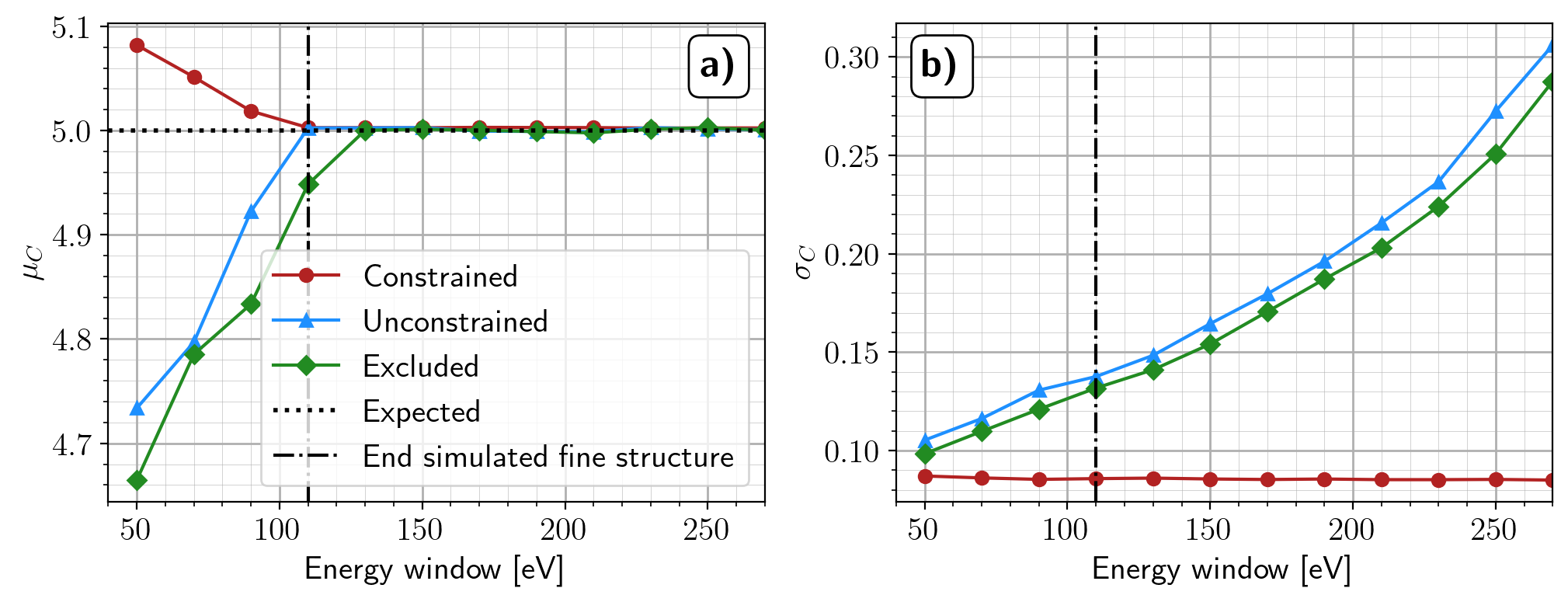}
\caption{(a) The average elemental abundance as a function of fine structure window for \unconstrained, \constrained \ and \excluded. The dotted horizontal line indicates the expectation value. The dash-dotted vertical line indicates the end of the simulated fine structure region. (b) The standard deviation of the elemental abundance as a function of fine structure window for the three methods. 
\label{fig:simulated_width}}
\end{figure}

\subsection{Simulation Study}
\label{sec:simstu}

\subsubsection{Influence of electron dose}

To illustrate the advantage of our method in terms of accuracy and precision we compare the \constrained, \unconstrained \ and \excluded \ method.  The fine structure windows of the three methods are set to $50$~eV. $\Delta E$ is set to the width of the zero-loss peak, and $110$ fine structure basis functions are used.

By using simulated data the methods' accuracy can be evaluated because the ground truth is known, which is difficult if not impossible for experimental EEL spectra.  Furthermore, as detailed in Sec.~\ref{sec:crlb}, this gives us the opportunity to set up the model such that the influence of the inequality constraints is negligible, resulting in meaningful CRBs.

The simulated core-loss spectrum consists of a background, a carbon K-edge atomic cross section, fine structure and low loss spectrum. The background is modelled as $cE^{-3}$,   The fine structure derived from an experimental diamond K-edge on which Eq.~\ref{eq:constrain} is enforced, and an inelastic mean free path of $0.3$ is chosen for the plasmons in the low loss spectrum. The full width half maximum of the zero-loss peak is 2~eV. 

In Fig.~\ref{fig:simulated}(a), the simulated spectrum is shown. For twenty noise levels, ranging from $450$ to $2.5 \times 10^8$ expected electrons, two thousand Poisson noise realizations each are simulated, and subsequently processed with the three proposed methods. The expected number of electrons was set through a scaling factor of the linear model parameters.

As can be seen from Fig.~\ref{fig:simulated}(b), no bias is observed for \constrained \ or \unconstrained. A bias of approximately $6\%$ is observed for \excluded. From Fig.~\ref{fig:simulated}(c) \constrained \ has best precision.  Preliminary results showed that for \excluded, precision can be traded off for bias through the exclusion window width, with a precision approaching that of \constrained \ at the cost of an even larger bias.

In Fig.~\ref{fig:simulated}(c), the precisions on the estimates from the noisy spectra are compared to the CRBs.  It is shown that of all estimation methods, \constrained \ has the lowest lower bound, and that all methods reach the CRB, which is remarkable for the \excluded \ method, as it is biased.

\subsubsection{Influence of fine structure width}

To illustrate the influence of the choice of fine structure energy window on the \constrained, \unconstrained \ and \excluded \ methods, the window is varied from 50~to~280~eV with the other components of the model kept constant. Again two thousand Poisson noise realizations of the model with true fine structure region of 110~eV (same as Fig.~\ref{fig:simulated}) are generated, and the average and standard deviations on the estimated elemental abundance are plotted in Fig.~\ref{fig:simulated_width}.

For \unconstrained, the precision is best for smallest fine structure energy windows, where the bias is largest. The bias can be reduced by extending the fine structure window, but this comes at the cost of a steadily degrading precision.  The \constrained \ method, on the other hand, shows a steady precision that is independent of fine structure window width, and a bias that is lower than \unconstrained's.  For both methods the bias vanishes once the fit encompasses the complete true fine structure window of 110~eV.  For \excluded \ the bias is highest, and the precision lies between that of the other two methods.

This illustrates the importance of selecting a proper fine structure window for \unconstrained \ and \excluded.  The optimal choice is not known \emph{a priori} and it is up to the user to make a judgement call in trading off bias versus precision.  No such issue is present with the \constrained \ method that combines lowest bias with best precision and an insensitivity to the width of the fine structure window. 

\begin{figure*}
\centering
\includegraphics[width=0.99\textwidth]{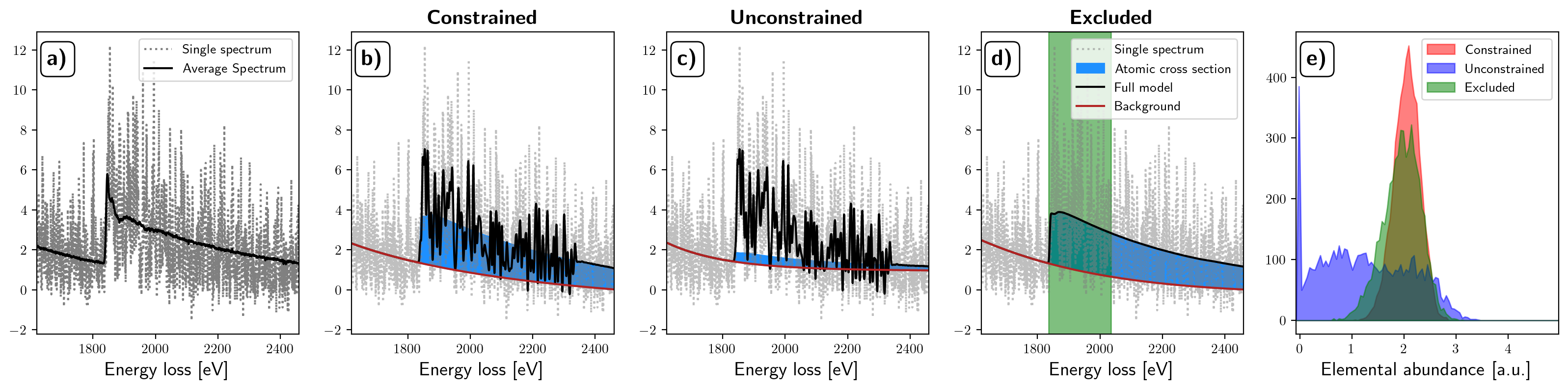}
\caption{(a) The solid and dotted line show the average of 5000 spectra and one single spectrum acquired on the \ce{Si} K-edge, respectively. (b) Visualization of a single fit with \constrained \ and the fitted atomic cross section in blue. (c) Same as (b) but with \unconstrained. (d) Same as (b) but with \excluded, the green region indicates the excluded $200$~eV fine structure window. (e) The histograms of the elemental abundance estimates of \unconstrained \ (blue), \constrained \ (red) and \excluded \ (green).
\label{fig:c_vs_nc}}
\end{figure*}

\begin{figure}
\centering
\includegraphics[width=0.99\linewidth]{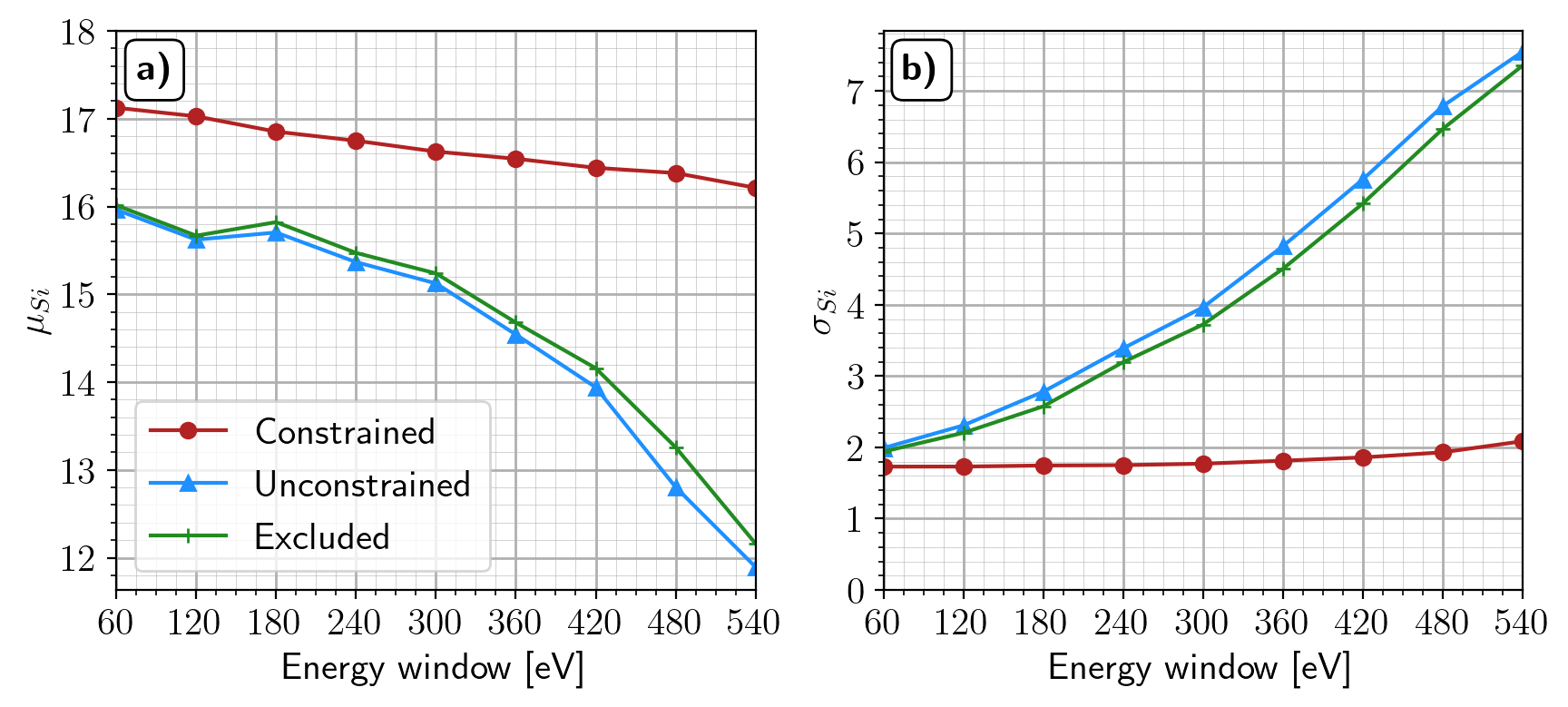}
\caption{The average Si (a) and standard deviation (b) as a function of energy width used for fitting of the fine structure.
\label{fig:window_l}}
\end{figure}

\subsection{Experimental Data}
\label{sec:expdat}

\unconstrained, \constrained \ and \excluded \ are tested on 5000 EEL spectra of the \ce{Si} K-edge, recorded on an amorphous \ce{Si_3N_4} sample.  All spectra stem from the same area, and a parallel beam was used to maximize illuminated area while minimizing beam damage and contamination.  Hence, the spectra are as identical as experimentally possible and differ only in their noise realization. In Fig.~\ref{fig:c_vs_nc}(a), the average and a typical single spectrum of the \ce{Si} K-edge are shown.

The acceleration voltage is $300$~keV, convergence and collection semi-angle are $0$~and~$50$~mrad, respectively. The energy loss ranges from $1620$~to~$2475$~eV, with a dispersion of $1$~eV.

The fitting model is described in Eq.~\ref{eq:model}, and has a single atomic cross section corresponding to the \ce{Si} K-edge located at $1839$~eV. For the fine structure window of \unconstrained \ and \constrained \ $\Delta E$ equals $4$~eV with $125$ basis functions, yielding a window of $500$~eV, while \excluded's fine structure window equals $200$~eV.

In Fig~\ref{fig:c_vs_nc}(b,c), the performance of \unconstrained \ and \constrained \ are illustrated on a single spectrum.  The former underestimates the atomic \ce{Si} K-edge, indicated by the blue area, and compensates with a too-intense fine structure.  The latter, however, does show a substantial atomic signal, combined with a fine structure that follows the equality constraint in Eq.~\ref{eq:constrain}, as can be gauged visually from the approximately zero fine structure integral.  The result of \excluded \ is illustrated in Fig.~\ref{fig:c_vs_nc}(d).

A statistically significant result is produced by analyzing all spectra and plotting the elemental abundances in a histogram in Fig.~\ref{fig:c_vs_nc}(e). \unconstrained \ produces a sharp peak at zero, caused by QP enforcing non-negativity.  The spreads produced by \constrained \ and \excluded \ are much smaller and the non-negativity constraints on the abundances do not come into play.

Next, sensitivity to the width of the fine structure windows is investigated, by varying them from $60$~to~$540$~eV in steps of $60$~eV.  The spectra are fitted with the three methods and the mean and standard deviation of the \ce{Si}-abundances are shown in Fig~\ref{fig:window_l}.  For \unconstrained \ and \constrained \ the width, $\Delta E$, of the basis functions is kept constant at $4$~eV. 

For \unconstrained \ and \excluded \ the mean abundances vary with $25$\% as a function of window width, and for \constrained \ with $6$\%.  The standard deviation increases by a factor of $4$ for \unconstrained \ and \excluded, while only a small increase of a factor of $1.2$ is observed for \constrained.  Moreover, also the absolute precision of \constrained \ is better over the entire investigated energy width.

It is thus shown that contrary to \unconstrained \ and \excluded, mean and precision of \constrained \ are hardly affected by the exact choice of fine structure window. This makes it a suitable method to reduce user interaction, as the needed parameter values can be set beforehand with hardly any negative consequences on the results.

\begin{figure}
\centering
\includegraphics[width=0.99\linewidth]{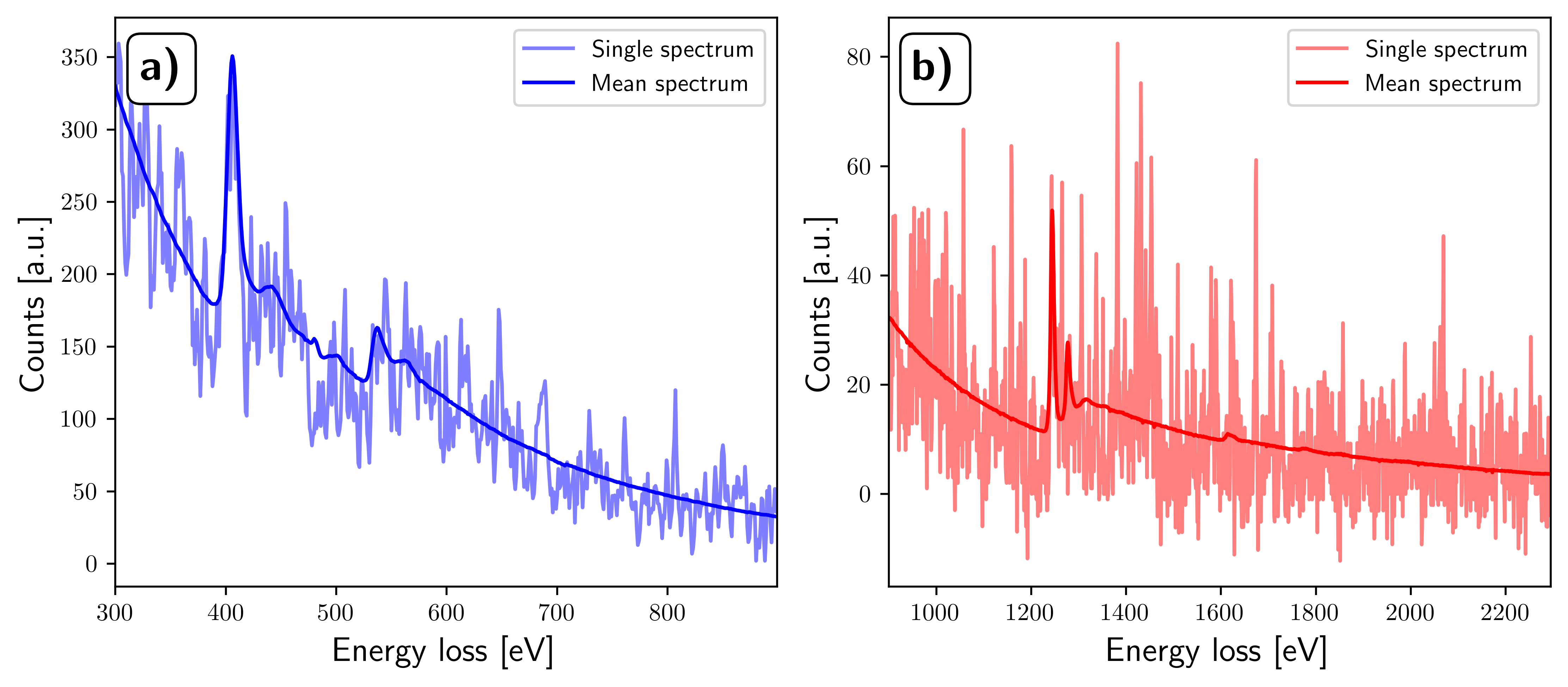}
\caption{The average and single EEL spectrum of \ce{TbScO3}.  The spectrum has been split in two parts for visualization purposes. \label{fig:SiN_s_m}}
\end{figure}

\begin{figure*}
\centering
\includegraphics[width=0.95\linewidth]{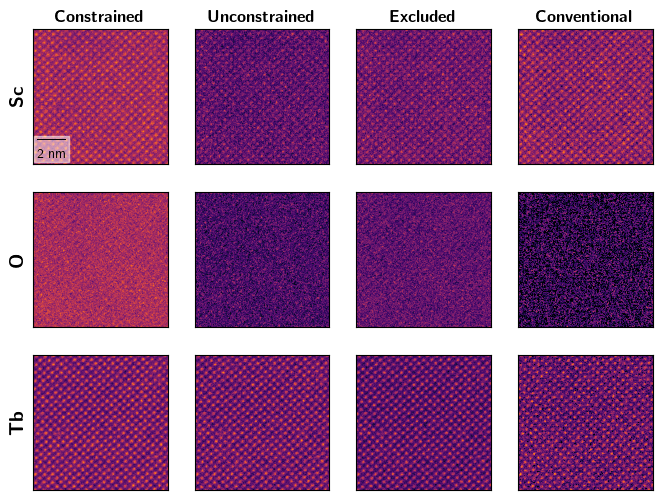}
\caption{The elemental maps of \ce{TbScO3} produced by the four investigated methods. 
\label{fig:TbScO_map}}
\end{figure*}

\begin{table}
\centering
\begin{tabular}{ |c|c|c|c| } 
 \hline
  & \ce{Sc} [\%] & \ce{O} [\%] & \ce{Tb} [\%] \\
  \hline
  \emph{Constr.} & $26.1 \pm 0.7$ & $56.4 \pm 0.7$ & $17.6 \pm 0.7$\\
  \emph{Unconstr.} & $26.6 \pm 1.9$ & $53.5 \pm 2.2$ & $19.9 \pm 1.0$\\
  \emph{Excl.} & $24.7 \pm 1.4$ & $57.0 \pm 1.6$ & $18.4 \pm 0.9$\\
  \emph{Conv.} & $26.3 \pm 2.4$ & $47.5 \pm 4.3$ & $26.2 \pm 2.4$\\
  \hline
\end{tabular}
 \caption{Average of the renormalized elemental abundances of the \ce{TbScO3}-sample estimated with \constrained, \unconstrained, \excluded \ and \conventional. The method \constrained \ attains lowest standard deviation across the elements.}
\label{table:tbsco3}
\end{table}

\begin{figure*}
\centering
\includegraphics[width=0.85\linewidth]{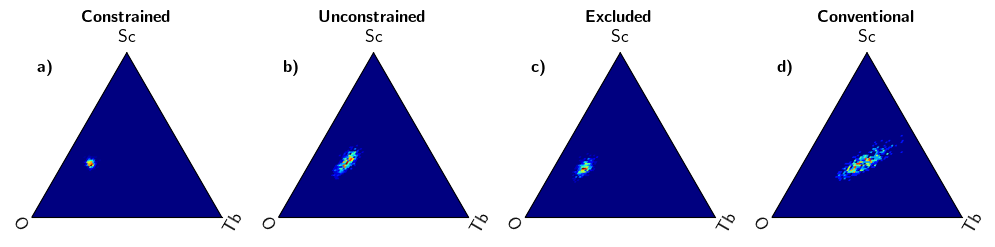}
\caption{The ternary plots \cite{mpltern} of the three elements for each method.  The results of \constrained \ have smallest spread and smallest correlation between elemental abundances.   
\label{fig:TbScO_ternary}}
\end{figure*}

\subsection{Elemental Mapping of \ce{TbScO3}}
\label{sec:elmatb}

The results of all four methods, \constrained, \unconstrained, \excluded \ and \conventional \ are compared by producing experimental elemental maps of a \ce{TbScO_3} crystal in the [001] direction.  The microscope was operated at $300$~kV with a convergence and collection semi-angle of respectively $25$~and~$30$~mrad. Since the incoming beam was convergent, the constraint on the fine structure from Eq.~\ref{eq:constraint_conv} were used.  The energy-loss ranges from $300$~to~$2300$~eV, with a dispersion of $1$~eV. In Fig. \ref{fig:SiN_s_m} the data is illustrated.  

The lengths of the fine structure windows of \excluded \ are set to $50$~eV, corresponding to the typical width of the ELNES region, except for the \ce{Tb} M$_{4,5}$-edge, where $120$~eV is used to encompass the strong white lines.

The elemental maps of the four methods are shown in Fig.~\ref{fig:TbScO_map}. Visual evaluation indicates that \constrained \ produces the best overall result, as it reveals atomic contrast in all three elemental maps, even for oxygen. Although \excluded \ attains an atomic contrast \ce{O}-map as well, its \ce{Sc}-map is much noisier. And while \conventional \ has excellent contrast for the \ce{Sc}-map, its \ce{O}-map is predominantly noise, because the \ce{O} K-edge is close to the tail of the \ce{Sc} L-edge (see Fig.~\ref{fig:SiN_s_m}(a)), thus strongly reducing the number of data points available for background estimation, leading to considerable extrapolation noise.

To achieve a more quantitative evaluation, we take the average of the elemental abundances \emph{within} each unit cell. These unit cells are identified by finding the positions of the \ce{Tb} atom columns via the StatStem open-source software \cite{debacker2016}, and using these as seeds to create Voronoi diagrams,~\cite{2020SciPy}, which are known to approximate well the unit cells in the sample, see \cite{rosenauer2011ultram}. 

The next step is to correct the average elemental abundances per unit cell by renormalizing the three estimates to a unit sum.  This step is needed to correct for thickness inhomogeneities.  We now assume further that the true elemental abundances are the same for each unit cell, and hence that the precision on the abundance estimates is given by the standard deviation \emph{between} unit cells.  

In Table.~\ref{table:tbsco3}, the averages and standard deviations of the elemental abundances thus obtained are shown.  It is clear that of all four methods \constrained \ yields the smallest standard deviation for the three elements. Furthermore, we cannot comment on the accuracy of the elemental abundance estimates because the dynamical scattering due to the sample's zone axis orientation modifies the apparent abundances.  Treating the associated channeling falls outside of the scope of this work and we refer to \cite{allen2015ultram} and \cite{lobato2016ultram} for a thorough treatment of the subject.

Further insight is gained with the ternary plots in Fig~\ref{fig:TbScO_ternary}, that provide a two-dimensional representation of the three-dimensional joint probability distribution of the \ce{Tb}, \ce{Sc} and \ce{O} abundances.  Not only is the distribution narrowest for \constrained, neither does it show the strong correlation between abundances from which the other three methods suffer.

\begin{figure*}
\centering
\includegraphics[width=0.71\linewidth]{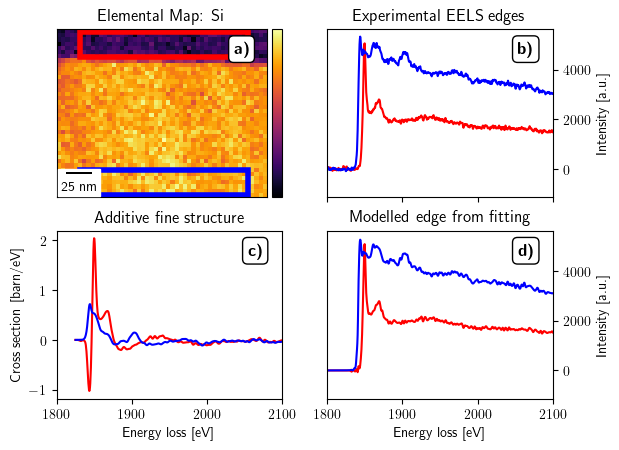}
\caption{(a)~\ce{Si} elemental map, the upper area contains \ce{SiO2} and the lower area \ce{Si}.  (b)~Normalized, background-subtracted spectra, averaged over the respective regions of interest in~(a). (c)~Additive fine structure, normalized with the atomic cross section.  (d)~The fitted fine structure and atomic cross section, note the resemblance with the experimentally measured edge in~(c).
\label{fig:finesi}}
\end{figure*}

\subsection{Fine structure of \ce{Si} and \ce{SiO2}}
\label{sec:finstrsisio}

In this section it is demonstrated how \constrained \ retrieves the fine structures associated with the \ce{Si} K-edge in a \ce{Si}-\ce{SiO2} sample, and that the result is a differentiator between the respective compounds and could hence be used to generate phase maps, for example with the aid of clustering algorithms like $k$-means clustering \cite{lloyd1982ieee}.  Note that since the low-loss spectrum is included in the model, the fine structure estimate is corrected for thickness effects and multiple scattering.

The acceleration voltage was $300$~kV, convergence and collection semi-angles were $20$~and~$40$~mrad, respectively, and the energy axis ranged from $1200$~to~$2400$~eV with a dispersion of $0.42$~eV.  
In Fig.~\ref{fig:finesi}.a) the \ce{Si} elemental map is shown and the regions used for further processing are indicated.  Due to the sample's uniform thickness, the \ce{SiO2} part has lowest intensity.  This property serves as ground-truth knowledge to distinguish between the two components and assures us that the stark difference in fine structure corresponds to a different electronic environment for \ce{Si}. Fig.~\ref{fig:finesi}.b) displays the averaged, normalized and background-subtracted spectra, and Fig.~\ref{fig:finesi}.c) the averaged additive fine structures.  Finally, Fig.~\ref{fig:finesi}.d) presents the fitted model which consists of the atomic cross section and the fitted fine structure.

\section{Discussion}
\label{sec:dis}

The analysis of the simulations in Sec. \ref{sec:simstu} showed that neither \constrained, \unconstrained \ nor \excluded \ suffer from bias once the fine structure window is large enough, and for small windows \constrained's bias is clearly lowest. Unfortunately, this behavior could not be reproduced with the experimental data in Sec. \ref{sec:expdat}: although the variability of the estimated average concentration was still lowest for \constrained, the estimation did not level off for large energy windows.  A possible cause might be a residual discrepancy between the atomic cross section model and reality.  While the used models are state-of-the-art, they do not include excitations to the bound unoccupied states, although formally that is a prerequisite for the validity of the Bethe-sum-rule-based equality constraint.

\section{Conclusion}
\label{sec:con}

In this work we showed how the Bethe sum rule leads to a linear equality constraint on the additive fine structure of EELS ionization edges.  Imposing this constraint on the fine structure estimate with quadratic programming improves the precision of the estimated elemental abundances.  Furthermore, the precision is proven close to optimal, as it approaches the CRB.  Compared to current estimation methods, be it the model-based approach with exclusion of the ELNES region or the conventional background extrapolation method, the accuracy improves as well.

Analysis beyond mere elemental abundance maps becomes possible because in the process the fine structure is estimated automatically and the result corrected for thickness and multiple scattering.  Since it reflects the crystal's unoccupied density of states, the fine structure provides information about the elements' chemical environment. Thus, the spectrum data is used to its full potential, as relevant information is retrieved instead of discarded.

Lastly, the results are insensitive to the exact values of the hyper-parameters, in particular the fine structure's range and the number of basis functions used to describe the fine structure.  Because of this behavior the estimation problem can be set up automatically and no other user input than the elemental content of the sample is needed as all other parameters, like for instance edge onsets, acceleration voltage or dispersion, are either tabulated or can be found in the spectra's metadata.  Hence an important source of user bias is eliminated.

\appendix

\section{Applying Bethe sum rule to cross sections}\label{app:theory}

In this section the constraint imposed on the fine structure in Eq.~\ref{eq:constrain} is derived. This is done by starting with the Bethe sum rule which states that when the system receives momentum $\bm q$, the transferred energy, summed over all internal modes of excitation, is the same as the energy transferred to $Z$ free electrons at rest: 
\begin{equation} \label{eq:sumrule}
    \sum_{n,l} \int_0^\infty \frac{df_{n,l}(E, \bm q)}{dE} dE  = Z, 
\end{equation}
where $Z$ is the total number of electrons in the  target atom and $df_{n,l}/dE$
is the generalized oscillator strength (GOS) of each sub-shell with principal quantum number of $n$ and angular quantum number of $l$, defined as,
\begin{equation}
    \frac{df_{n,l}(E, \bm{q})}{dE} =\frac{\Delta E}{R} \frac{\sum_{f} |\langle f |  e^{i\bm{q}\cdot \bm{r}} | i \rangle |^2}{(qa_0)^2} \delta(\Delta E - E),
\end{equation}
where $| i \rangle$ is the initial state, $| f \rangle$ is the final state of target orbital electron and $\Delta E$ is the transferred energy between $| i \rangle$ and $| f \rangle$. Note that Eq.~\ref{eq:sumrule} assumes non-relativistic wave functions; we refer to  \cite{cohen2004aqc} for an extension to the relativistic case.

Since only the inner-shell excitation's are of interest in this work, the sum rule is rewritten for a single sub-shell as, 
\begin{equation} \label{eq:inner_sr}
    \int_0^\infty \frac{df_{n,l}(E, \bm q)}{dE} dE  = Z_{n,l},
\end{equation}
with $Z_{n,l}$ the number of electrons in the sub-shell; for example, for K-edges $Z_{1,0} = 2$, and for L$_{2,3}$-edges $Z_{2,1} = 6$.  Here, following \cite{inokuti1971romp}, the wave functions for the different sub-shells are approximated as independent.

We now follow \cite{crozier1989ultram}. The cross section of the sub-shell $\sigma_{in}(E)$ associated with the quantum numbers $n$ and $l$ is given by:
\begin{equation}\label{eq:gos_to_cross}
    \sigma_{in}(E) = \frac{4 a_0^2R^2}{E T} \int_0^{\beta} \frac{2\pi \theta}{\theta^2 + \theta_E(E)^2} \frac{df_{n,l}(E, \bm q)}{dE} d\theta,
\end{equation}
where $a_0$ is the Bohr radius, $\beta$ the collection angle, $R$ the Rydberg energy, $T$ the effective incident energy and $\theta_E$ the characteristic angle given by, 
\begin{equation}
    \theta_E(E) = \frac{E}{2\gamma T}.
\end{equation}
For small collections angles, the GOS is approximated by a constant and brought outside of the integral, yielding:
\begin{equation}
    \sigma_{in}(E) = \frac{4 \pi a_0^2R^2}{E T} \log \Big(1+\frac{\beta^2} {\theta_E(E)^2}\Big)\frac{df_{n,l}(E, 0)}{dE}.
\end{equation}
Moving all the factors dependent on $E$ to the left, except for the GOS, and integrating over the energy loss yields,
\begin{equation} \label{eq:constant}
\begin{aligned}
    \int_0^{\infty} & \frac{E \cdot \sigma_{in}(E)}{\log \Big(1+\frac{\beta^2} {\theta_E(E)^2}\Big)} dE \\ 
    & = \frac{4 \pi a_0^2R^2}{T} \int_0^{\infty} \frac{df_{n,l}(E, 0)}{dE} = \text{constant}. 
\end{aligned}
\end{equation}
The left hand side is constant and holds true for arbitrary cross sections, independent of the system's internal states, so also for atomic cross sections for free atoms, $\sigma_A$.  The latter are theoretically calculated and known well --- see \cite{leapman1980jcp, segger2022zenodo, zhang2024zenodo, zhang2024relativistic} --- and hence are used as a reference to arrive at the result in Eq.~\ref{eq:constrain}.

\section{Linear and Multiplicative Fine Structure} \label{app:1}

Here follows a rigorous derivation of the equivalence of a multiplicative and linear fine structure model. The cross section in the multiplicative approach is given as,
\begin{equation}
    \sigma(E) = \sum a_i\cdot g_i(E) \cdot \sigma_A(E),
\end{equation}
and for the additive approach as,
\begin{equation}
    \sigma(E) = \sum b_i \cdot g_i(E) + \sigma_A(E).
\end{equation}
The functions $g$ are orthogonal fine structure basis set.

The first step is to equate both expressions, multiply by $g_i(E)$ and sum over the entire energy range.
\begin{eqnarray}
\begin{aligned}    
    a_i & \sum_{E_n}  g_i(E_n)^2 \sigma_A(E_n) \\ 
    & = b_i \sum_{E_n}  g_i(E_n)^2 + \sum_{E_n} \sigma_A(E_n) f_i(E_n)
\end{aligned} 
\\
\begin{aligned}
    b_i = & \frac{1}{\sum_{E_n}  g_i(E_n)^2} \bigg( a_i \sum_{E_n}  g_i(E_n)^2 \sigma_A(E_n) \\
    & - \sum_{E_n} \sigma_A(E_n) g_i(E_n) \bigg)
\end{aligned}
\end{eqnarray}
By using Dirac delta functions as a basis set, $g_i(E)=\delta(E-E_i)$, the following result is obtained
\begin{equation}
    b_i = a_i\sigma_A(E_i) - \sigma_A(E_i) = \sigma_A(E_i) (a_i - 1), \label{eq:fs_trafo}
\end{equation}
which provides a convenient transformation between the multiplicative and the linear approach.

\section{Equality constraints for Convergent Incoming Electron Beam} 
\label{app:2}

The derivation in Appendix \ref{app:theory} assumes that the incoming electrons are described by a plane wave. However, in many experimental setups, a convergent beam is used. In this appendix, the equality constraints for the convergent incoming beam will be derived. In the work by \cite{kohl1985}, the effective cross section is derived and following result is obtained: 
\begin{equation}
\begin{aligned}
    \sigma_{in}(E) = \frac{4 a_0^2R^2}{E T} & \frac{df_{in}(E, 0)}{dE} \\ 
    & \times \int_0^{\beta+\alpha} F(\theta) \frac{2\pi \theta}{\theta^2 + \theta_E(E)^2}  d\theta
\end{aligned}
\end{equation}
This is similar to Eq.~\ref{eq:gos_to_cross}, but an extra factor $F(\theta)$ is added which is a correction factor that takes the convergence angle of the incoming beam into account and is given by the expression:
\begin{equation}
    F(\theta) =
    \begin{cases}
    \sbullet \ \min(1, \beta^2 / \alpha^2), \quad \text{for } 0\leq \theta \leq |\alpha-\beta|, \\
    \sbullet \ (1 / \pi) \Big[ \arccos(x) + \\
    (\beta^2 / \alpha^2) \arccos(y) - \\
    (1 / 2\alpha^2) \sqrt{4\alpha^2\beta^2 + (\alpha^2 + \beta^2 - \theta^2)^2} \Big], \\ 
    \quad \quad \text{for } |\alpha - \beta| < \theta \leq \alpha + \beta,
    \end{cases}
\end{equation}
where 
\begin{equation}\label{eq:conv_const_cb}
    \begin{split}
        x = \frac{\alpha^2 + \theta^2 - \beta^2}{2\alpha \theta}, \\
        y = \frac{\beta^2 + \theta^2 - \alpha^2}{2\beta \theta}.
    \end{split}
\end{equation}

A derivation similar to Appendix~\ref{app:theory} yields 
\begin{equation}\label{eq:constraint_conv}
\sum_i b_i \int_0^{E_{max}} \frac{E g_i(E)}{\int_0^{\beta+\alpha} F(\theta) \frac{2\pi \theta}{\theta^2 + \theta_E(E)^2}  d\theta} dE = 0, 
\end{equation}
where the integral in the denominator of is evaluated numerically instead of analytically.

\section{Covariance matrix for maximum likelihood estimation with linear equality constraints}
\label{app:crlb}

Assuming the measurements, $g_i$, for bin $i$ are random draws from a Poisson distribution with expectation value $f_i$, where $f$ is a model depending on the $N$ parameters in the vector $p$, constrained minimization of the negative log-likelihood is written as,
\begin{eqnarray}
    \begin{aligned}
   \arg & \min_p \sum_i f_i(p) - g_i \ln (f_i(p)),\\ 
   & \text{subject to } \sum_{j \in \C} w_j p_j = 0,
   \end{aligned}
\end{eqnarray}
where $\C$ is the set of indices indicating the elements of $p$ that partake in the linear equality constraint, $w_j$ for $j \in \C$ are the known weights, for example given by the integral in the summands of Eq.~\ref{eq:constrain} and $w_k = 0$ for $k \notin \C$. 
 The problem is solved by finding the stationary points of the Lagrangian,
\begin{eqnarray}
    \L(p; \lambda) = \sum_i f_i(p) - g_i \ln(f_i(p)) + \lambda \sum_{j \in \C} w_j p_j,
\end{eqnarray}
as a function of the $N$ parameters in $p$ and the multiplier $\lambda$.  In other words, for the estimation problem the system of $N+1$ equations,
\begin{eqnarray}
    F_{\ell}  & = & \frac{\partial \L}{\partial p_{\ell}} = 0, \text{ for } \ell = 1, \ldots, N,\\
    F_{N+1} & = & \frac{\partial \L}{\partial \lambda} = 0,
\end{eqnarray}
must be solved. A new vector $q = (p, \lambda)^T$ containing all parameters including $\lambda$ is defined.  The explicit expressions for the functions $F$ are,
\begin{eqnarray}
    \frac{\partial \L}{\partial p_k} &=&  \sum_i \left(1 - \frac{g_i}{f_i} \right) \frac{\partial f_i}{\partial p_k}, \text{ for } k \notin \C, \\
    \frac{\partial \L}{\partial p_j} & = & \sum_i \left( 1 - \frac{g_i}{f_i} \right) \frac{\partial f_i}{\partial p_j} + \lambda w_j, \text{ for } j \in \C, \label{eq:lambda} \\
    \frac{\partial \L}{\partial \lambda} & = & \sum_{j \in \C} w_j p_j.
\end{eqnarray}

Fortunately, an expression for the covariance matrix can be retrieved without explicitly solving the system. Consider that the covariance matrix of the parameters $q$ is given by the standard expression for the propagation of errors,
\begin{eqnarray}
    \cov_q = \left. \frac{\partial q}{\partial g} \cov_g \frac{\partial q}{\partial g}^T \right\vert_{g = f},
\end{eqnarray}
which, following \cite{barlow1991_43}, is evaluated in the true values for $g_i$, namely $f_i$. The covariance matrix $\cov_g$ is given by $\text{diag}(f)$, and for the partial derivatives the implicit function theorem is invoked:
\begin{eqnarray}
    \frac{\partial q}{\partial g}  =  \left(  \frac{\partial F}{\partial q} \right)^{-1}  \frac{\partial F}{\partial g}.
\end{eqnarray}
Working this out yields,
\begin{eqnarray}
    \left. \frac{\partial F_{\ell}}{\partial q_k} \right\vert_{g = f} & = &   \sum_i \frac{1}{f_i}  \frac{\partial f_i}{\partial q_{\ell}} \frac{\partial f_i}{\partial q_k} \text{ for } \ell, k = 1 \ldots N, \\
     \left. \frac{\partial F_k}{\partial \lambda} \right\vert_{g = f} & = & 0 \text{ for } k \notin \C \text{ or } k = N+1, \\
     \left. \frac{\partial F_j}{\partial \lambda} \right\vert_{g = f} & = & w_j \text{ for } j \in \C.
\end{eqnarray}
For the derivatives with respect to $g_i$ we have,
\begin{eqnarray}
    \left. \frac{\partial F_{\ell}}{\partial g_i} \right\vert_{g_i = f_i} & = & \frac{-1}{f_i}  \frac{\partial f_i}{\partial q_{\ell}} \text{ for } \ell = 1 \ldots N, \\
    \left. \frac{\partial F_{N+1}}{\partial g_i} \right\vert_{g_i = f_i}  & = & 0,
\end{eqnarray}
which yields,
\begin{equation}
\begin{split}
    & \left( \frac{\partial F}{\partial g}  \text{diag}(f)  \frac{\partial F}{\partial g}^T \right)_{j,k} \\ 
    & = \sum_i \frac{1}{f_i}  \frac{\partial f_i}{\partial q_j} \frac{\partial f_i}{\partial q_k} (1 - \delta_{j, N+1}) (1 - \delta_{k, N+1}), \\
    & \text{ for } j, k = 1 \cdots N+1.
\end{split}
\end{equation}

Finally the covariance matrix, $\cov_q$, is obtained and represented in a more visual form as,
\begin{strip}
\begin{eqnarray}
    \begin{pmatrix}
     & & & & & & \vline & w_1 \\
     & & & & & & \vline &  \\
     \multicolumn{6}{c}{\scalebox{1.41}{$\sum_i \frac{1}{f_i}  \frac{\partial f_i}{\partial p_j} \frac{\partial f_i}{\partial p_k}$}} & \vline & \vdots \\
     & & & & & & \vline &  \\
     & & & & & &  \vline &  w_N \\
     \hline
     w_1 &  &  & \hdots &  & w_N & \vline & 0
    \end{pmatrix}^{-1}
    \begin{pmatrix}
     & & & & & \vline & 0 \\
     & & & & & \vline &   \\
     \multicolumn{5}{c}{\scalebox{1.41}{$\sum_i \frac{1}{f_i}  \frac{\partial f_i}{\partial p_j} \frac{\partial f_i}{\partial p_k}$}} & \vline & \vdots \\
     & & & & & \vline &   \\
     & & & & &  \vline &  0 \\
     \hline
     0 &  & \hdots &  & 0 & \vline & 0
    \end{pmatrix}
    \begin{pmatrix}
     & & & & & & \vline & w_1 \\
     & & & & & & \vline &  \\
     \multicolumn{6}{c}{\scalebox{1.41}{$\sum_i \frac{1}{f_i}  \frac{\partial f_i}{\partial p_j} \frac{\partial f_i}{\partial p_k}$}} & \vline & \vdots \\
     & & & & & & \vline &  \\
     & & & & & &  \vline &  w_N \\
     \hline
     w_1 &  &  & \hdots &  & w_N & \vline & 0
    \end{pmatrix}^{-1^T}, \label{eq:crlb_long}
\end{eqnarray}
\end{strip}

where we recall that the vector $p$ contains the $N$ model parameters, and that the weights $w_k$ equal zero for the parameters not partaking in the equality constraint, \emph{i.e.} for $k \notin \mathcal{C}$.

The extension to more than one equality constraint is trivial.

Eq.~\ref{eq:crlb_long} provides the covariance matrix of $q$, i.e. of all parameters including $\lambda$, although in the current context the multiplier is a mere help-variable and neither its exact value nor precision is of interest to us. Numerical tests indicate that the covariance matrix of the model parameters, $p$, is given by the first $N$ rows and $N$ columns of the first factor in Eq.~\ref{eq:crlb_long}, thus yielding the final result in Eq.~\ref{eq:crlb_short}.

It is interesting to note that in the absence of the equality constraint, the last row and column of the individual factors are not present, so that the expression reduces to the conventional, constraint-free form of the Cram\'er-Rao lower bound under Poisson noise:
\begin{eqnarray}
     \left( \sum_i \frac{1}{f_i}  \frac{\partial f_i}{\partial p} \frac{\partial f_i}{\partial p} \right) ^{-1}. \label{eq:crlb}
\end{eqnarray}

The linearity of the constraints comes into play via Eq.~\ref{eq:lambda}, since derivatives of this function with respect to the parameters in $p$ and with respect to $\lambda$ both remove the multiplier from the problem.

\FloatBarrier

\section*{Data availability}

Access to the source code for \constrained \ and the \ce{TbScO3} dataset is provided in \cite{verbeeck2024github} and \cite{jannis2024zenodo}, respectively.

\section*{Acknowledgments}

This project has received funding from the ECSEL Joint Undertaking (JU) under grant agreement No 875999. The JU receives support from the European Union’s Horizon 2020 research and innovation programme and Netherlands, Belgium, Germany, France, Austria, Hungary, United Kingdom, Romania, Israel.

\small

\bibliographystyle{apalike}
\bibliography{refs}    

\begin{thebibliography}{}

\bibitem[Allen et~al., 2015]{allen2015ultram}
Allen, L., {D'Alfonso}, A., and Findlay, S. (2015).
\newblock Modelling the inelastic scattering of fast electrons.
\newblock {\em Ultramicroscopy}, 151:11--22.

\bibitem[Barlow, 1991]{barlow1991_43}
Barlow, R.~J. (1991).
\newblock {\em A Guide to the Use of Statistical Methods in the Physical
  Sciences}, chapter 4.3.
\newblock John Wiley \& Sons, Chichester.

\bibitem[Bertoni and Verbeeck, 2008]{verbeeck2008ultram}
Bertoni, G. and Verbeeck, J. (2008).
\newblock Accuracy and precision in model based {EELS} quantification.
\newblock {\em Ultramicroscopy}, 108:782--790.

\bibitem[Cohen, 2004]{cohen2004aqc}
Cohen, S.~M. (2004).
\newblock Aspects of relativistic sum rules.
\newblock In {\em Theory of the Interaction of Swift Ions with Matter. Part 2},
  volume~46 of {\em Advances in Quantum Chemistry}, pages 241--265. Academic
  Press.

\bibitem[Cram\'er, 1946]{cramer1946book}
Cram\'er, H. (1946).
\newblock {\em Mathematical Methods of Statistics}.
\newblock Princeton Univ. Press, Princeton, NJ.

\bibitem[Crozier and Egerton, 1989]{crozier1989ultram}
Crozier, P. and Egerton, R. (1989).
\newblock Mass-thickness determination by bethe-sum-rule normalization of the
  electron energy-loss spectrum.
\newblock {\em Ultramicroscopy}, 27(1):9--18.

\bibitem[Cueva et~al., 2012]{cueva2012}
Cueva, P., Hovden, R., Mundy, J.~A., Xin, H.~L., and Muller, D.~A. (2012).
\newblock Data processing for atomic resolution electron energy loss
  spectroscopy.
\newblock {\em Microscopy and Microanalysis}, 18(4):667–675.

\bibitem[{De Backer} et~al., 2016]{debacker2016}
{De Backer}, A., {van den Bos}, K., {Van den Broek}, W., Sijbers, J., and {Van
  Aert}, S. (2016).
\newblock Statstem: An efficient approach for accurate and precise model-based
  quantification of atomic resolution electron microscopy images.
\newblock {\em Ultramicroscopy}, 171:104--116.

\bibitem[{d}en Dekker et~al., 2013]{dendekker2013ultram}
{d}en Dekker, A.~J., Gonnissen, J., {D}e {B}acker, A., Sijbers, J., and {V}an
  {A}ert, S. (2013).
\newblock Estimation of unknown structure parameters from high-resolution
  {(S)TEM} images: What are the limits?
\newblock {\em Ultramicroscopy}, 134:34--43.

\bibitem[Egerton and Malac, 2002]{egerton2002ultram}
Egerton, R. and Malac, M. (2002).
\newblock Improved background-fitting algorithms for ionization edges in
  electron energy-loss spectra.
\newblock {\em Ultramicroscopy}, 92(2):47--56.

\bibitem[Egerton, 2011a]{egerton1996_44}
Egerton, R.~F. (2011a).
\newblock {\em {E}lectron {E}nergy-{L}oss {S}pectroscopy in the {E}lectron
  {M}icroscope}, chapter 4.4.
\newblock Springer, New York, Dordrecht, Heidelberg, London, 3$^{\text{rd}}$
  edition.

\bibitem[Egerton, 2011b]{egerton1996_38}
Egerton, R.~F. (2011b).
\newblock {\em {E}lectron {E}nergy-{L}oss {S}pectroscopy in the {E}lectron
  {M}icroscope}, chapter 3.8.
\newblock Springer, New York, Dordrecht, Heidelberg, London, 3$^{\text{rd}}$
  edition.

\bibitem[Goldfarb and Idnani, 1983]{goldfarb1983}
Goldfarb, D. and Idnani, A. (1983).
\newblock A numerically stable dual method for solving strictly convex
  quadratic programs.
\newblock {\em Mathematical Programming}, 27:1--33.

\bibitem[Ikeda, 2024]{mpltern}
Ikeda, Y. (2024).
\newblock mpltern.

\bibitem[Inokuti, 1971]{inokuti1971romp}
Inokuti, M. (1971).
\newblock Inelastic collisions of fast charged particles with atoms and
  molecules---{T}he {B}ethe theory revisited.
\newblock {\em Reviews of Modern Physics}, 43:297--347.

\bibitem[Jannis et~al., 2024a]{jannis2024zenodo}
Jannis, D., Van~den Broek, W., and Verbeeck, J. (2024a).
\newblock Electron energy loss dataset on {TbScO}$_3$.
\newblock \url{https://doi.org/10.5281/zenodo.14005449}.

\bibitem[Jannis et~al., 2024b]{verbeeck2024github}
Jannis, D., Zhang, Z., Verbeeck, J., and Annys, A. (2024b).
\newblock py{EELSMODEL}.
\newblock \url{https://github.com/joverbee/pyEELSMODEL}.

\bibitem[Kohl, 1985]{kohl1985}
Kohl, H. (1985).
\newblock A simple procedure for evaluating effective scattering cross-sections
  in stem.
\newblock {\em Ultramicroscopy}, 16(2):265--268.

\bibitem[Leapman et~al., 1980]{leapman1980jcp}
Leapman, R.~D., Rez, P., and Mayers, D.~F. (1980).
\newblock {{K}, {L}, and {M} shell generalized oscillator strengths and
  ionization cross sections for fast electron collisions}.
\newblock {\em The Journal of Chemical Physics}, 72(2):1232--1243.

\bibitem[Lehmann and Casella, 2006]{lehmann2006book}
Lehmann, E.~L. and Casella, G. (2006).
\newblock {\em Theory of point estimation}.
\newblock Springer Science \& Business Media.

\bibitem[Liu and Brown, 1987]{liu1987jom}
Liu, D.-R. and Brown, L.~M. (1987).
\newblock Influence of some practical factors on background extrapolation in
  eels quantification.
\newblock {\em Journal of Microscopy}, 147(1):37--49.

\bibitem[Lloyd, 1982]{lloyd1982ieee}
Lloyd, S.~P. (1982).
\newblock Least square quantization in {PCM}.
\newblock {\em IEEE Transactions on Information Theory}, 28:129--137.

\bibitem[Lobato et~al., 2016]{lobato2016ultram}
Lobato, I., Van~Aert, S., and Verbeeck, J. (2016).
\newblock Progress and new advances in simulating electron microscopy datasets
  using {MULTEM}.
\newblock {\em Ultramicroscopy}, 168:17--27.

\bibitem[McGibbon, 2021]{quadprog}
McGibbon, R.~T. (2021).
\newblock quadprog 0.1.11.
\newblock \url{https://github.com/quadprog/quadprog}.
\newblock Released: Nov 27, 2021.

\bibitem[Moore, 2010]{moore2010thesis}
Moore, T.~J. (2010).
\newblock {\em A Theory of Cram\'er-Rao Bounds for Constrained Parametric
  Models'}.
\newblock Phd thesis, University of Maryland, College Park, MD, 20742.

\bibitem[Nocedal and Wright, 1999]{nocedal1999}
Nocedal, J. and Wright, S.~J. (1999).
\newblock {\em {N}umerical {O}ptimization}.
\newblock {S}pringer {S}eries in {O}perations {R}esearch. Springer, New York,
  Dordrecht, Heidelberg, London, 1$^{\text{st}}$ edition.

\bibitem[Rao, 1945]{rao1945bul}
Rao, C.~R. (1945).
\newblock Information and the accuracy attainable in the estimation of
  statistical parameters.
\newblock {\em Bulletin of the Calcutta Mathematical Society}, 37:81--89.

\bibitem[Rez, 2004]{rez2004}
Rez, P. (2004).
\newblock {\em Energy Loss Fine Structure}, chapter~4, pages 97--126.
\newblock John Wiley \& Sons, Ltd.

\bibitem[Rosenauer et~al., 2011]{rosenauer2011ultram}
Rosenauer, A., Mehrtens, T., Müller, K., Gries, K., Schowalter, M., {Venkata
  Satyam}, P., Bley, S., Tessarek, C., Hommel, D., Sebald, K., Seyfried, M.,
  Gutowski, J., Avramescu, A., Engl, K., and Lutgen, S. (2011).
\newblock Composition mapping in ingan by scanning transmission electron
  microscopy.
\newblock {\em Ultramicroscopy}, 111(8):1316--1327.

\bibitem[Segger et~al., 2022]{segger2022zenodo}
Segger, L., Guzzinati, G., and Kohl, H. (2022).
\newblock {A set of Generalised Oscillator Strengths for the simulation of EELS
  spectra}.

\bibitem[Su and Zeitler, 1993]{su1993prb}
Su, D.~S. and Zeitler, E. (1993).
\newblock Background problem in electron-energy-loss spectroscopy.
\newblock {\em Phys. Rev. B}, 47:14734--14740.

\bibitem[{v}an~{d}en {B}os and {d}en {D}ekker, 2001]{vandenbos2001hawkes}
{v}an~{d}en {B}os, A. and {d}en {D}ekker, A.~J. (2001).
\newblock {\em Resolution Reconsidered---Conventional Approaches and an
  Alternative}, pages 242--360.
\newblock Academic Press, United States.

\bibitem[Van~den Broek et~al., 2023]{vandenbroek2023ultram}
Van~den Broek, W., Jannis, D., and Verbeeck, J. (2023).
\newblock Convexity constraints on linear background models for electron
  energy-loss spectra.
\newblock {\em Ultramicroscopy}, 254:113830.

\bibitem[Van~den Broek et~al., 2019]{vandenbroek2019ieee}
Van~den Broek, W., Reed, B.~W., Béché, A., Velazco, A., Verbeeck, J., and
  Koch, C.~T. (2019).
\newblock Various compressed sensing setups evaluated against {S}hannon
  sampling under constraint of constant illumination.
\newblock {\em IEEE Transactions on Computational Imaging}, 5(3):502--514.

\bibitem[Verbeeck and {Van Aert}, 2004]{verbeeck_2004}
Verbeeck, J. and {Van Aert}, S. (2004).
\newblock Model based quantification of eels spectra.
\newblock {\em Ultramicroscopy}, 101(2):207--224.

\bibitem[Verbeeck et~al., 2006]{verbeeck2006ultram}
Verbeeck, J., Van~Aert, S., and Bertoni, G. (2006).
\newblock Model-based quantification of {EELS} spectra: Including the fine
  structure.
\newblock {\em Ultramicroscopy}, 106:976--980.

\bibitem[Virtanen et~al., 2020]{2020SciPy}
Virtanen, P., Gommers, R., Oliphant, T.~E., Haberland, M., Reddy, T.,
  Cournapeau, D., Burovski, E., Peterson, P., Weckesser, W., Bright, J., {van
  der Walt}, S.~J., Brett, M., Wilson, J., Millman, K.~J., Mayorov, N., Nelson,
  A. R.~J., Jones, E., Kern, R., Larson, E., Carey, C.~J., Polat, {\.I}., Feng,
  Y., Moore, E.~W., {VanderPlas}, J., Laxalde, D., Perktold, J., Cimrman, R.,
  Henriksen, I., Quintero, E.~A., Harris, C.~R., Archibald, A.~M., Ribeiro,
  A.~H., Pedregosa, F., {van Mulbregt}, P., and {SciPy 1.0 Contributors}
  (2020).
\newblock {{SciPy} 1.0: Fundamental Algorithms for Scientific Computing in
  Python}.
\newblock {\em Nature Methods}, 17:261--272.

\bibitem[Zhang et~al., 2024a]{zhang2024zenodo}
Zhang, Z., Lobato, I., Brown, H., Lamoen, D., Jannis, D., Verbeeck, J., Aert,
  S.~V., and Nellist, P. (2024a).
\newblock {Generalised oscillator strength for core-shell electron excitation
  by fast electrons based on Dirac solutions}.

\bibitem[Zhang et~al., 2024b]{zhang2024relativistic}
Zhang, Z., Lobato, I., Brown, H., Lamoen, D., Jannis, D., Verbeeck, J., Aert,
  S.~V., and Nellist, P.~D. (2024b).
\newblock Relativistic {EELS} scattering cross-sections for microanalysis based
  on {D}irac solutions.

\end{thebibliography}

\end{document}